\documentclass[onecollarge,final]{svjour2}
\usepackage[latin1]{inputenc}
\usepackage[english]{babel}
\usepackage[namelimits]{amsmath}
\usepackage{amssymb}
\usepackage{amsmath}
\usepackage{color}
\usepackage{dcolumn}

\usepackage{subfig}
\usepackage{graphicx}
\begin{document}
\title{A statistical Test of Walrasian Equilibrium by Means of Complex Networks Theory}

\author{Leonardo Bargigli \and Stefano Viaggiu \and Andrea Lionetto}


\institute{L. Bargigli \at Dipartimento di Scienze per l'Economia e l'Impresa, Universit\`{a} di Firenze,
              Via delle Pandette 32, 50127 Firenze (Italy), 
              \email{leonardo.bargigli@unifi.it}
              \and
	      S.Viaggiu \at Dipartimento di Matematica, Universit\'a `Tor Vergata', Via della Ricerca Scientifica, 1, Rome, Italy 00133,
	      \email{viaggiu@axp.mat.uniroma2.it}
	      \and A. Lionetto \at Dipartimento di Fisica, Universit\'a `Tor Vergata',Via della Ricerca Scientifica, 1, Rome, Italy 00133, \email {lionetto@roma2.infn.it}
}

\date{Received: date / Accepted: date}

\date{\today}
\maketitle
\begin{abstract}
We represent an exchange economy in terms of
statistical ensembles for complex networks by introducing the concept of market configuration.
This is defined as a sequence of nonnegative discrete random variables $\{w_{ij}\}$ describing the flow of a given commodity from agent $i$ to agent $j$. This sequence can be arranged in a nonnegative matrix $W$ which we can regard as the representation of a weighted and directed network or digraph $G$.
Our main result consists in showing that general equilibrium theory imposes highly restrictive conditions upon market configurations, which are in most cases not fulfilled by real markets. An explicit example with reference to the e-MID interbank credit market is provided.
\end{abstract}
\maketitle
\section{Introduction}
The first attempt to draw a link between classical thermodynamics and economics, due to \cite{2}, was based on the parallel between entropy maximization and utility or profit maximization. Under this perspective, economic agents are treated like thermodynamic macroscopic subsystems composing a larger macroscopic system which is the market \cite{3b}. This view, although consistent with the supposed optimizing behavior of economic agents, is at odds with the statistical physics microfoundation of thermodynamics, according to which macroscopic equilibrium arises from the (random) behavior of micro units. Statistical physics introduces a clear separation of the microscopic and macroscopic levels which is lost in the proposed economic parallelism.

A recent stream of literature employs the techniques of statistical mechanics under the hypothesis that in a limited period of time an economic system may behave as though in a quasi-equilibrium state 
\cite{1,b4,b2,b3,3,3a,15,4,5,6,3c,4d,44d,1K}.
In these models we have homogeneous agents randomly exchanging money with a constraint on the total amount of money $M$ in the system.
The homogeneity of economic agents is a consequence of the assumption of uniform (in physical terms, isotropic) random exchange, since the latter hypothesis makes agents symmetric with respect to the probability distribution of their monetary holdings.
In this paper we attempt to take a step forward by describing an economic system with heterogeneous agents. We resort to complex networks theory in order to introduce the key notion of market configuration. This is defined as a sequence of nonnegative discrete random variables $\{w_{ij}\}$ describing the flow of a given commodity from $i$ to $j$. This sequence can be arranged in a nonnegative matrix $W$ which we can regard as the representation of a weighted and directed network or digraph $G$.
This is by no means a completely new idea. In the wake of the financial crisis, many scholars have directed their efforts to the application of network theory to real markets, and in particular to credit markets\footnote{Recent reviews are \cite{upper2011} and \cite{chinazzi2013}.}.

Although the main preoccupation of these works is the empirical analysis of contagion processes, some scholars have focused also on the theoretical side, trying to develop network models providing a faithful representation of markets \cite{Marsili2012}.
Some of these works \cite{fagiolo2013} employ the statistical mechanics approach of \cite{2p,3g}, where the authors consider a set of network realizations (ensemble) and then impose constraints on the expectation value of a given set of graph observables $\{x_i\}$ with respect to the ensemble. This approach, as we see in the next section, lends itself to a very natural interpretation in terms of economic equilibrium.
After a detailed description and construction of the statistical ensembles for complex newtorks, our attention will be focused on
the application of the statistical mechanics techniques in order to study real markets and the role played by the thermodynamic
temperature $T$. As an explicit example, in order to test WE against a real system, we use the e-MID interbank credit
transactions data. The e-MID interbank credit market represents a centralized and fully transparent platform, thus complying with the basic framework of WE theory. In the period under consideration (January 2005), moreover, rates where stable and with a very limited dispersion, so that it is reasonable to assume market equilibrium with a common price that was virtually equal to the policy rate. The main result is to show that WE is testable and indeed too restrictive to describe e-MID market configurations.

The structure of the paper is the following. In section \ref{sec: prel} we discuss the main ideas to build the ensembles of complex networks. In section \ref{sec: ens} we build the ensembles. In section \ref{sec: appl} we illustrate the role played by $T$ in terms of statistical uncertainty over market states. The main result of this section consists in showing that general equilibrium theory imposes highly restrictive conditions upon market configurations. In section 5 we further analyze the parallel between thermodynamic and economic equilibrium as addressed in the economic literature. Section 6 is devoted to some conclusions.

\section{Preliminaries} \label{sec: prel}

In microeconomic theory an exchange economy is a system of $N$ consumers, initially endowed with $\omega_i$ units of the available $M$ commodities and with preferences over alternative consumption vectors $x_i \in B_i(p)$, where $B_i(p)$ (the price-dependent \textit{budget set}) is a set of affordable consumption vectors for the consumer $i$, given $\omega_i$ and $p$. In this economy, trade occurs whenever agents prefer a consumption vector $x_i \neq \omega_i$. The excess demand function $z_i = x_i - \omega_i$ describes the trading plan of each agent.

A \textit{Walrasian} equilibrium (WE) occurs when the market-clearing condition is satisfied for some non negative price vector $p^*$:

\begin{equation}\sum_i z_i(p^*) = 0 \label{wal_eq}
\end{equation}

and at the same time the consumption vector of each individual is the preferred one $x^*_i$ in her own budget set. The existence of the equilibrium price vector $p^*$, which can be proved invoking Katutani's fixed-point theorem, is regarded as a fundamental achievement of microeconomic theory \cite{arrowdebreu54} \footnote{ Arrow-Debreu and Radner equilibria extend WE by introducing the notion of contingent commodities, i.e. of commodities whose delivery is conditional on a realized state of the world. The extension of our framework to these equilibria is obtained by associating an independent network ensemble to each state $s$.}.

In order to connect WE and network theory, we resort to the notion of market configuration defined in the previous section. With this definition in mind, we introduce the following notations: $x_i = \sum_j w_{ji}$ is the final allocation for agent $i$ in the market under consideration \footnote{For simplicity we omit market indices and refer in the following to the single commodity case. The extension to multiple markets is straightforward. In fact, we can treat multiple markets at economic equilibrium as a set of statistically independent networks \cite{bianconi13}.}, and the excess demand on this market is defined as follows

\begin{equation}
 z_i = \sum_{j \neq i}\left( w_{ji} -  w_{ij}\right) = x_i - \sum_{j} w_{ij}. 
\end{equation}

Thus we obtain that $\omega_i = \sum_j w_{ij}$, where it is understood that $w_{ii}$ stands for the fraction of the initial endowment which is not exchanged on the market. 
For each market configuration and each agent $i = 1,2,\dots,N$, we can make the following identifications:

\begin{align}\omega_i & = w_i^{out} \\
x_i & = w_i^{in} \\
z_i & = w_i^{in} - w_i^{out}
\end{align}

where $w_i^{out} \equiv \sum_j w_{ij}$,$w_i^{in} \equiv \sum_i w_{ij}$ are respectively the out-strength and in-strength of node $i$.
Since in this way $x$ and $\omega$ are identified with network observables, namely the out- and in-strength sequences $w^{out},w^{in}$ of $G$, we can apply in this context the approach of \cite{2p}, which allows to generate ensembles of random graphs $\mathcal{G}$ displaying a set of desired average properties $\left\lbrace x_{i}\right\rbrace$.

In practice, these authors consider a set of graph configurations $G\in \mathcal{G}$ imposing the constraint that the expectation value of graph observables $\{x_i\}$ with respect to the ensemble probability distribution $P(G)$ is equal to some arbitrary value $\bar{x}_i$:
\begin{eqnarray}\label{eq: gen_constr}
& &\langle x_i \rangle = \sum_{G}\,x_i(G)\,P(G) = \bar{x}_i 
\end{eqnarray}
$P(G)$ is obtained maximizing the Gibbs entropy
together with the observables and normalization constraints.
As a result, they obtain the distribution
\begin{eqnarray}
& & P(G)=\frac{e^{-H(G)}}{Z},\;\;Z=\sum_{G\in \mathcal{G}} e^{-H(G)}, \nonumber\\
& & H(G)=\sum_{i} {\theta}_i x_i(G), \label{p3}
\end{eqnarray}
where $\{{\theta}_{i}\}$ are Lagrange multipliers.

The rationale for adopting entropy maximization in the economic context is the following. According to Walrasian theory, economic agents in an exchange economy care only about final allocations. Thus they are indifferent between market configurations which yield the same final allocations, i.e. they would be satisfied to pick any of them at random. Thanks to entropy maximization, the frequency of market configurations in $\mathcal{G}$ reflects this indifference.

The degree and strength sequences are the most common observables chosen as constraints in this problem.
By strength of a node $i$ in a weighted symmetric network we define the sum $w_i = \sum_{j} w_{ij}$. If $W$ is asymmetric, i.e. $G$ is directed, we need to distinguish between the out-strength $w_i^{out}$ and in-strength $w_i^{in}$, as explained above. The degree of a node $i$, instead, is defined over the binary adjacency matrix $A$, representing a binary network, as the sum $k_i = \sum_{j} a_{ij}$. If $A$ is asymmetric, again we need to distinguish between the out-degree and in-degree of the node $i$.
\cite{2p} provide a solution when the constraints are represented either by the degree or by the strength sequences. They arrive at the following Hamiltonian for a binary undirected network with fixed expected degree sequence:

\begin{equation}
H=\sum_{i}{k}_{i}{\theta}_{i}.
\label{pp1}
\end{equation}

Using the definition of $k_i$, eq. (\ref{pp1}) may be rewritten as follows

\begin{equation}
\label{p8}
H = \sum_{i<j}{\epsilon}_{ij}{a}_{ij},
\end{equation}

where $\epsilon_{ij} = \theta_i + \theta_j$.
For a weighted undirected network they obtain instead
\begin{equation}
\label{p9}
H = \sum_{i<j}{\epsilon}_{ij}{w}_{ij}.
\end{equation}
Starting from the expressions (\ref{p8})-(\ref{p9}), they derive the analogue of the quantum Fermi-Dirac and Bose-Einstein distributions respectively for $\left<a_{ij}\right>$ and $\left<w_{ij}\right>$.
By inspection it is easy to see that the ${w}_{ij}$ and $a_{ij}$ in eqs. (\ref{p8}) and (\ref{p9}) are indeed equivalent to the occupation numbers included in the Hamiltonian of the Fermi and Bose ideal gases respectively, while the $\epsilon_{ij}$ are equivalent to the energy levels occurring in the same Hamiltonian.
This observation justifies the interpretation, made by the authors, according to which links are the equivalent of particles in network ensembles.
In economic terms, these links / particles are commodity units while the energies $\epsilon_{ij}$ are very naturally interpreted as shadow prices \cite{foley94}.
It's straightforward to get to the conclusion that, if $w_i^{out}$ and $w_i^{in}$ are the observables of choice and if the constraints (\ref{eq: gen_constr}) take the form
\begin{align} \label{eq: walr_cons1}
\langle w^{out} \rangle & = x^*(p^*) \\
\langle w^{in} \rangle & = \omega  \label{eq: walr_cons2}
\end{align}
then the average market configuration $\langle G \rangle$ will be a WE. In fact the market clearing condition is automatically satisfied since $\sum_i w^{out}_i \equiv \sum_i w_i^{in}$. In order to take into account the small statistical fluctuations of strengths in the ensemble, we introduce the notion of \textit{statistical Walrasian equilibrium} as a network ensemble for which the following holds:
\begin{align}
w^{out} & \approx x^*(p^*) \\
 w^{in} & \approx \omega
\end{align}
where $w^{out},w^{in}$ are the observed out- and in-strength sequences of a random member $G$ of the ensemble $\mathcal{G}$.
From the viewpoint of physical theory, we should note the difference of the distribution (\ref{p3}) with respect to the usual Boltzmann-Gibbs distribution, namely that the former is coincident with the latter only for $T=1$.

Since $T$ is related to the energy constraint $H(G) = \bar{E}$, we see that energy is not explicitly conserved in the network ensemble, but its conservation is a consequence of the constraints over observables. In fact, the energies in (\ref{p8}) or (\ref{p9}) can be rescaled in order to obtain a common factor $T$ by making the substitution $\epsilon_{ij} = \epsilon_{ij}'/T$. Of course, it is always possible to specify the energy levels in such a way as to verify the constraints (\ref{eq: gen_constr}) for $T = 1$ or some other fixed value, by making the substitution $\epsilon_{ij} = \epsilon_{ij}'/T_0$ for some fixed $T_0$. At the same time, if $T \neq T_0$ the constraints cannot be satisfied. Thus we see that  the Park \& Newman and the ideal gas model of statistical physics are not necessarily consistent. 
In the following, we derive network ensembles along the standard lines of physical theory, thereby recovering the factor $\beta = 1/T$ of statistical physics which is the multiplier for the energy constraint. The main motivation for proceeding in this way is that we wish to estimate $T$ from real market data in order to derive a statistical test for the null hypothesis that the observed market configuration represents a WE  (see sec. \ref{sec: appl}).  In this way we define a hierarchy of increasingly restrictive models. In the first place we have the general model (GM) characterized by the following constraint

\begin{equation}
H(G) = \bar{E}
\end{equation}

for the generic Hamiltonian

\begin{equation} \label{eq: ham}
H(G) = \sum_{ij} \epsilon_{ij} w_{ij}
\end{equation}

with arbitrary energy levels $\left\lbrace \epsilon_{ij} \right\rbrace$.
The corresponding multiplier is $\beta$. By imposing the further constraints on the strength sequences
$w^{out},w^{in}$ with Lagrange multipliers
$\{{\lambda}_i\}, \{{\theta}_i\}$, we obtain that ${\epsilon}_{ij}={\lambda}_i+{\theta}_j$ in (\ref{eq: ham}). This is the Park \& Newman model (PNM). We recall that the constraints over strengths sequences are satisfied only for a fixed temperature $T_0$. Under the further assumption that
the market-clearing condition is obtained with the equilibrium values $(x^*,p^*)$, a Walrasian equilibrium model (WEM) arises. In this case the   constraints over strength sequences are provided by the equations (\ref{eq: walr_cons1})-(\ref{eq: walr_cons2}). We can represent the situation depicted above
in the following manner:
\begin{eqnarray}
& & \mathcal{G}_{GM}\;\supset\;\;\mathcal{G}_{PNM}\;\supset\;\mathcal{G}_{WEM}, \label{ss1}\\
& & \{H\}\rightarrow \{H, w_i^{out}, w_i^{in}\}\rightarrow\nonumber\\
& &\rightarrow \{H, w_i^{out},w_i^{in}, x^*(p^*) \}\label{ss2}\\
& & \{\beta\}\rightarrow \{\beta, {\lambda}_i, {\theta}_i\}\rightarrow \{\beta, {\lambda}_i, {\theta}_i, p^*\}\label{ss3},
\end{eqnarray}
where (\ref{ss2}) refers to the conserved quantities and (\ref{ss3}) refers to the resolutive values of the three models $\mathcal{G}_{GM}, \mathcal{G}_{PNM},\mathcal{G}_{WEM}$ in the order represented in (\ref{ss1}).

The scheme (\ref{ss1}) clearly indicates the possibility to represent a given WEM in terms of the more general (less restrictive)
GM model once the energies are suitably specified as ${\epsilon}_{ij}={\lambda}_i+{\theta}_j$, the multipliers $\left\lbrace  \lambda_i \right\rbrace, \left\lbrace  \theta_i \right\rbrace $ satisfy the constraints (\ref{eq: walr_cons1})-(\ref{eq: walr_cons2}) and $T = T_0$. 
In the following we identify the equivalent of the volume $V$ of physical systems for networks with the degrees of freedom of the network itself. These are given by the number of points of the bidimensional lattice formed by the couples of nodes $(i,j)$, i.e. $V = O(N^2)$. For example,
for directed graphs we have $V=N(N-1)$,  for undirected ones $V=N(N-1)/2$
while for directed graphs with self loops $V = N^2$. Following the economic interpretation, we will generally refer to the last specification. Then we see that the Hamiltonian (\ref{eq: ham}) is a sum over discrete spatial coordinates $x,y = 1, \dots,N$ contained in a square of side $(N - 1)\,d$. The lattice distance $d$ is a convenience parameter which we might let go to zero in case we wish to switch from discrete to continuous coordinates.

According with the usual interpretation \cite{2p,3g}, the number of links, that we denote from now on with $L$, is the formal analogue of the number particles in the usual physical systems. In the economic interpretation, the fluctuations of $L$, introduced with the grandcanonical ensemble, are a consequence of uncertainty over the available commodities under different states of the world,
since $L = \sum_i \omega_i$.\\

\section{Statistical ensembles} \label{sec: ens}

\subsection{Microcanonical ensemble} \label{sec: micro}

Let's suppose that agents are certain about the state of the world $\bar{s}$, so that their endowments are exactly determined $\omega_i = \omega_i(\bar{s})$. Then, provided that their preferences are well behaved and that they know the market clearing price $p^*$, they can choose their optimal consumption $x^*$ and the market will display with certainty an ``energy'' level $E = \sum_i \left(\lambda_i x^*_i + \theta_i \omega_i\right)$. Here the $\lambda_i$, $\theta_i$ are nothing more than multipliers for the following constraints:
\begin{align}
w_i^{out} & = \omega_i(\bar{s}) \label{eq: constr1}\\
w_i^{in} & = x_i^*(\bar{s}) \label{eq: constr2}
\end{align}
In general there are many market configurations which are consistent with these constraints. Since agents have no reason to prefer one of this states over the other, the latter can be considered as equally likely.\\
In this ensemble, the ``energy'' $H(G)=E$ is conserved and the number of links or commodity units $L$ and of agents $N$ are fixed too. These are the \textit{thermodynamic constraints } of the system. 
The ordinary microcanonical ensemble $\mathcal{G}_{GM}$ describes the statistical equilibrium of a system with arbitrary energy levels. In this ensemble, all configurations which comply with the thermodynamic constraints are equally likely. The WE ensemble $\mathcal{G}_{WEM}$ requires additionally that the \textit{economic constraints} (\ref{eq: constr1})-(\ref{eq: constr2}) are satisfied. Among the possible energy levels, we define as Walrasian energy levels $\epsilon^*_{ij} = \lambda_i + \theta_i$ those levels that verify the economic constraints.
By considering the thermodynamic constraints only, as usual, we can define $\Gamma(E)$ as  the total number of configurations, at fixed $V = O(N^2)$ and $L$, calculated at the surface $H(G)=E=const.$ of constant energy. Hence, the entropy $S$ can be defined in the usual way:
\begin{equation}
S(E,V,L)=\ln\Gamma(E,V,L) = \ln \left ( \sum_{G} W(G) \right)
\label{cqrt}
\end{equation} 
 The sum is limited to those network configurations which satisfy $H(G) = E$, and $W(G)$ counts the number of link micro-states corresponding to a given network configuration. Then the usual Bose-Einstein, Fermi-Dirac and Boltzmann distributions are derived from a suitable specification of $W(G)$.These coincide, as expected, with the average link values which are derived from the grandcanonical ensembles (see the appendix).
\subsection{Canonical ensemble}
Now let's suppose that the probability distribution of states is peaked around a given state $\bar{s}$ and that $x^*$, $\omega$ and $p^*$ are continuous in $s$. In this case the conditions for a statistical WE, specified in sec. \ref{sec: prel}, are fulfilled. Agents trade according to the observed $s$, but the probability of observing large deviations from WE, as defined in the previous section, is small.
Inspired by this economic view,
the derivation of the canonical ensemble follows the lines present in \cite{1,huang}.
Firstly we introduce $G_1$ and $G_2$ as sub-networks of an isolated network $G$ of fixed energy $E$ with $E_1 \ll E_2$. Then we assume that $H(G) \simeq H(G_1) + H(G_2)$, which means that the states $(i,j)$, defined between couples of nodes $i,j$ belonging respectively to $G_1$ and $G_2$, are empty, i.e. that links or commodity units cannot flow from one subnetwork to the other. Thus the two systems can be treated as mutually independent. Under the assumption that
$E$ is conserved, we have
\begin{equation}
\Gamma(E)\simeq {\Gamma}_1(E_1){\Gamma}_2(E_2=E-E_1)
\label{app8}
\end{equation}
In relation to (\ref{cqrt}),
the following thermodynamical relations hold:
\begin{eqnarray}
& &dS=\frac{\partial S}{\partial E}dE+\frac{\partial S}{\partial V}dV,\label{app2}\\
& &{\left(\frac{\partial S}{\partial V}\right)}_{E,L}=\frac{P}{T},\;\;
{\left(\frac{\partial S}{\partial E}\right)}_{V,L}=\frac{1}{T}, \label{app3}
\end{eqnarray}
By performing a Taylor expansion with $E_1 \ll E$, we get
\begin{eqnarray}
& &S(E-E_1)=S(E)+\frac{\partial S(E)}{\partial E}(-E_1)+o(1)=\nonumber\\
& &=S(E)-\frac{E_1}{T}+o(1).
\label{c2}
\end{eqnarray}
where the last equality in (\ref{c2}) follows from (\ref{app3}).
Since we are interested in the behavior of the subsystem $1$ independently from the reservoir $2$
we can write, remembering that $E = H(G)$,
\begin{equation}
Z=\sum_{G}e^{-\frac{H(G)}{T}},\;\;\;P(G)=C e^{-\frac{H(G)}{T}},\;\;C=\frac{1}{Z}.
\label{app9}
\end{equation}
As stated above, differently from (\ref{p3}), the standard factor $\beta$ appears explicitly
in (\ref{app9}). 
The statistical equilibrium described by the canonical ensemble is also a WE when
the last lines in (\ref{ss2}) and (\ref{ss3}) are satisfied.
\subsection{Grand canonical ensemble}
Since the Park and Newman distribution is a special case of the standard Gibbs distribution, the grand canonical extension may be accomplished along the standard lines\footnote{We observe that Park and Newman don't derive explicitly the grand canonical partition function, although they use it implicitly to solve their model. Even if they don't introduce the chemical potential, their results are correct since, as underlined in \cite{3g}, the latter can be always absorbed in the energy terms $\epsilon_{ij}$.}. Here it is essential to 
stress again that the number of agents $N$ is related to the volume of the network,
i.e. $V=O(N^2)$ is the equivalent of the volume $V$ of physical systems.
We suppose that $G_1$ and $G_2$ are still (quasi) isolated components of a larger system G characterized by the
Gibbs distribution (\ref{app9}), but now links are allowed to move between the two volumes $V_1$ and $V_2$.
The partition function $Z$ of $G$ is
\begin{eqnarray}
& & Z_G(L,V,T)  = \sum_G e^{ -\frac{H(G)}{T} }=\\
& &= \sum_{L_1 = 0}^{L} \sum_{G_1} e^{ -\frac{H(G_1)}{T} }
\sum_{G_2} e^{ -\frac{H(G_2)}{T} }.\nonumber
\end{eqnarray}
Following the usual derivation in statistical mechanics \cite{huang} we can write the partition function as follows
\begin{equation}
Z=e^{-\frac{F}{T}}.
\label{app10}
\end{equation}
where $F$ is the analogue of the free Helmholtz energy:
\begin{eqnarray}
& &F=E-TS,
\end{eqnarray}
With the help of (\ref{app9}) and (\ref{app10}), we set $\mu= {\left(\frac{\partial F}{\partial L}\right)}_{V,T}$,
$P=-{\left(\frac{\partial F}{\partial V}\right)}_{L,T} $ and
introduce the fugacity $z = \exp \left( \frac{\mu}{T}\right)$. Then, after a Taylor expansion,
we obtain the grand canonical partition function:
\begin{equation}
\mathcal{Q}(z,V,T) \equiv \sum_{L = 0}^{\infty}z^L Z_G(L,V,T),
\label{app11}
\end{equation}
where the subscript '1' has been dropped in (\ref{app11}). Moreover, the following standard relationships still hold:
\begin{align}
\log \mathcal{Q} & = \dfrac{PV}{T} \label{gingol}\\
 L & = z \, \, \dfrac{\partial}{\partial\,z} \log \mathcal{Q}\label{bel},
\end{align}
For a Boltzmann gas, these relationships allow to derive immediately the equation of state $\frac{PV}{T} = L$. 
In fact, in this case we have $ L = \log \mathcal{Q}$ (see appendix).
We see that the ratio $\frac{P}{T}$ coincides with an observable quantity, namely the link density of the network.
\subsection{Reversible transformations}
Like in thermodynamics, the equation of state prescribes which combinations of state variables $(T,P,V,L)$ are necessarily associated with a maximal entropy value, i.e. with macroscopic equilibrium, for the system under consideration. It is then theoretically conceivable to move the system along these combinations, i.e. without leaving macroscopic equilibrium.
In our case $P$ is not determined independently, so the equation of state is always satisfied.
By analogy, we may nevertheless think of a transformation changing the vertex number $N$ (i.e. $V$). As an example, if vertices represent a system of banks, then this can be accomplished by a merger of two or more banks or by the failure of some bank.
By eq. (\ref{eq: ham}), instead, energy exchange can be obtained by a change in the number of links $L$. A variation of $L$, which represents a net inflow or outflow of commodities, has a clear interpretation in terms of open economic systems.
Then we are naturally brought to wonder which are the economic implications of reversible transformations, i.e. if WE is preserved when $(T,V,L)$ are changed. In order to answer the question, we need to examine the effect of a change of the independent state variables $(T,V,L)$ over the constraints (\ref{eq: constr1})-(\ref{eq: constr2}). It's clear that any change breaks down the constraints (\ref{eq: constr1})-(\ref{eq: constr2}), thus it's impossible to obtain reversible transformation preserving economic equilibrium.

\section{Real Markets}\label{sec: appl}

In this section we apply the statistical ensembles introduced above, and in particular the  grand canonical one, to study the role of temperature within the economic interpretation of network ensembles\footnote{ In the literature, the concept of graph temperature has been introduced for the first time in \cite{2006cond.mat..6805G} within a grand canonical ensemble. Moreover, in \cite{kalinka}  the temperature of a complex network is studied in terms of the clustering properties of the graph.}.
The well known Sonnenschein-Mantel-Debreu (SMD) theorem states that for any collection of price vectors $\left\lbrace p^i\right\rbrace$, $i = 1, \dots,M$ and matrices of price effects  $\left\lbrace \nabla z(p^i)\right\rbrace$, there is an exchange economy for which these price vectors are equilibrium price vectors, with price effects at equilibrium given by the corresponding matrices. The standard interpretation of this result is that WE cannot be falsified by the empirical evidence provided by price effects. Without contradicting the SMD theorem, we show that WE is instead highly restrictive with respect to market configurations. Then real market configurations may falsify WE, turning the latter into a testable theory.
 
To start with, the grand canonical partition function defined in (\ref{app11}) can be rewritten as:
\begin{equation}
Z =\sum_{G \in \mathcal{G}} \exp(\mu L_G-H_G)/T,
\end{equation}
where $L_G$ is the number links in $G$.  The probability of a graph $G$ is given by:
\begin{equation}
 P_G=\frac{1}{Z}\exp(\mu L_G-H_G)/T
\label{eq:grandcanprob}
\end{equation}
The energy (hamiltonian) function of $G$ is given by eq.~(\ref{eq: ham}):
\begin{equation}
 H_G = \sum_{ij}\epsilon_{ij} {w}_{ij}
\end{equation}
In the following we extend the results obtained by
\cite{2006cond.mat..6805G} for the two limiting cases $T\rightarrow\infty$ and $T\rightarrow 0$.
We begin by studying the limit for $T\rightarrow\infty$:
\begin{equation}
 \lim_{T\rightarrow \infty}P_G=Z^{-1} = |\mathcal{G}|^{-1}
\end{equation}
i.e. all the graphs in the ensemble have the same probability. We can regard this limit as equivalent to the Sonnenschein-Mantel-Debreu theorem. In fact, no market configuration $G$ can be excluded with high probability from the ensemble $\mathcal{G}_{WEM}$ in this limit.
 
In the low temperature limit $T\rightarrow 0$ we obtain instead a delta distribution centered on the most likely configuration $G^*$:
\begin{align}
\lim_{T\rightarrow 0} P_{G^*} & = 1 \\
\lim_{T\rightarrow 0} P_{G \neq G^*} & = 0
\end{align}
$G^*$ is obtained from the following maximization problem:
\begin{equation}
G^* = \text{argmax} \,\, \left( \mu L_G - H_G \right )
\end{equation}
\begin{figure}[htbp]
\begin{center}
\subfloat[]{\includegraphics[scale=0.5]{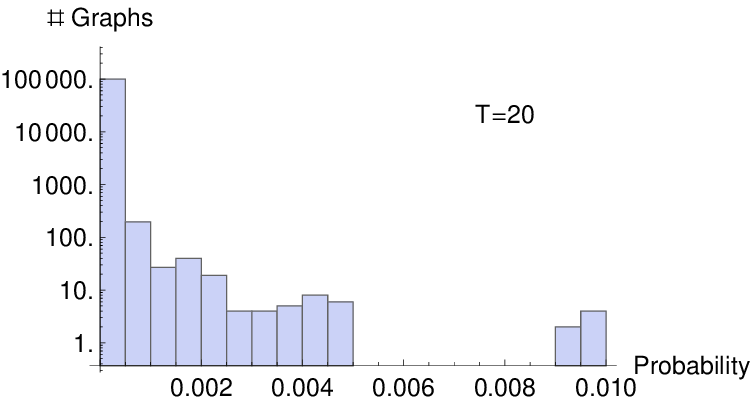}}\\
\subfloat[]{\includegraphics[scale=0.5]{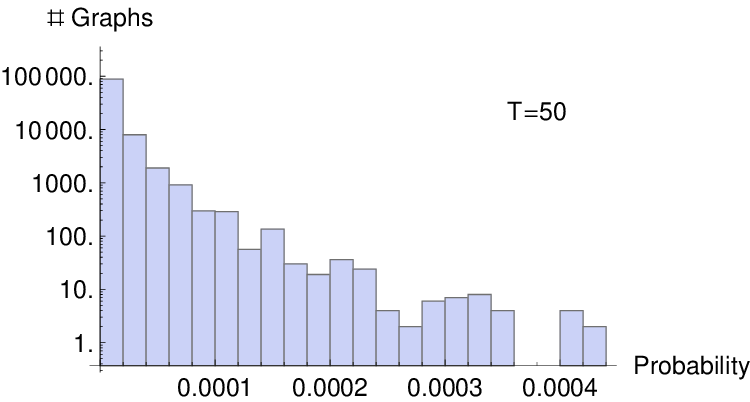}}\\
\subfloat[]{\includegraphics[scale=0.5]{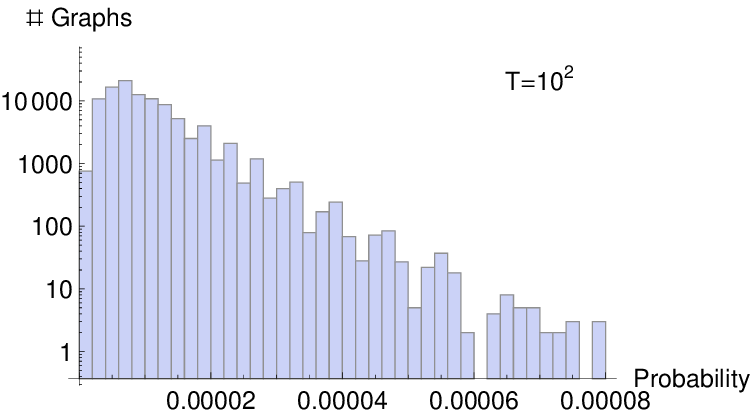}}\\
\subfloat[]{\includegraphics[scale=0.5]{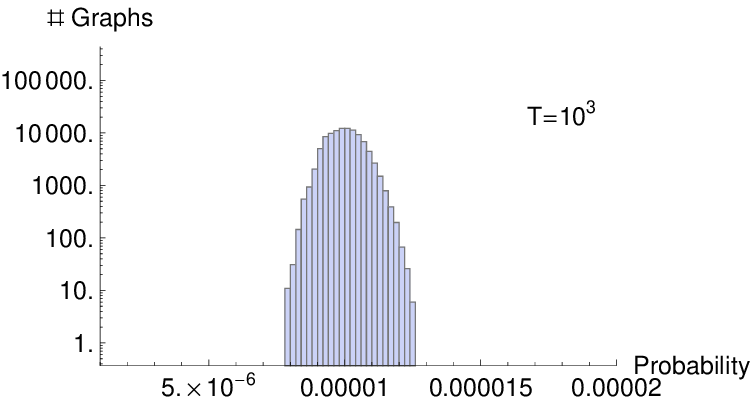}}
\caption{Number of graphs with given probability for $T=20,\,50,\,100\,1000$.}
\label{fig:T10T50T100}
\end{center}
\end{figure}
Let us now see what happens for $T$ interpolating between these two exact results. We will rely on some numerical random graph sampling. The sampling dimension is $10^5$. We assume $\epsilon_{ij}$ gaussian iid variables with $\bar{\epsilon}_{ij}=1$ and standard deviation $\sigma=0.5$. The results for $T=20,\,50,\,100$ are encoded in Fig. \ref{fig:T10T50T100} showing for a fixed temperature the number of graphs as a function of their associated probability. The probability is the grandcanonical one written in eq.~(\ref{eq:grandcanprob}). It is possible to see that in the first two cases a small subset of graphs has a higher likelihood, while almost all graphs have negligible probability. For $T=10^2$ the probability starts spreading more evenly among the graphs. This behavior is even clearer taking into account higher temperature values (panel (d)).
For these higher temperatures the ``collapse'' of all the graphs towards the probability $10^{-5}$, allowed by the finite sampling, is well rendered.

From this exercise we see that the value of $T$ is of crucial importance for the predictive power of the Walrasian model: the smaller $T$, the larger the number of market configurations that can be excluded w.h.p. from the $\mathcal{G}_{WEM}$ ensemble. Actually one does not have to take into account in general the whole set of states but only the fraction with non negligible probability which, as we have seen, crucially depends on $T$.

\begin{figure}[htbp]
\begin{center}
\includegraphics[scale=0.3]{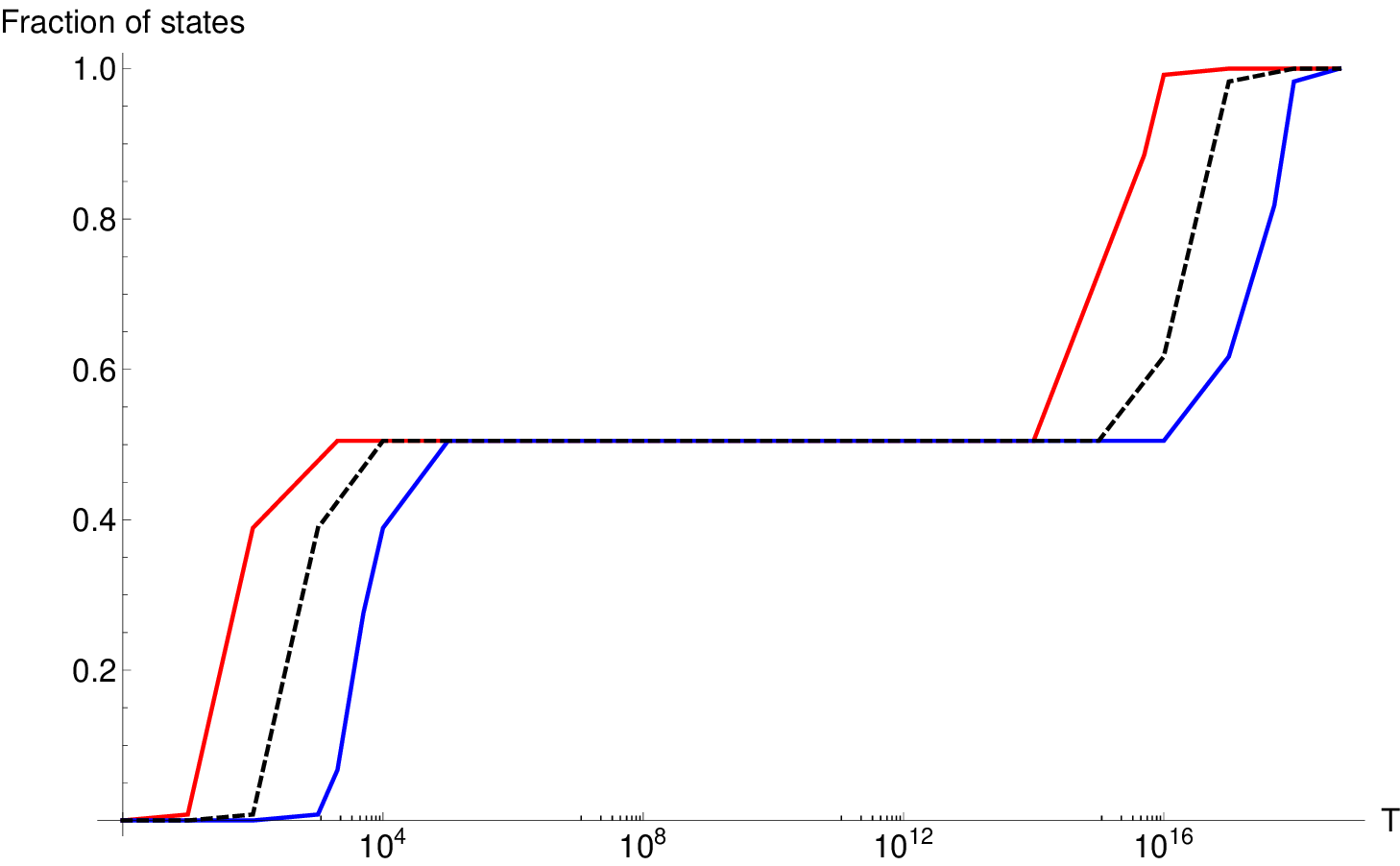}
\caption{Fraction of states (sample size $S=10^4$) above a given probability as a function of $T$. The red curve has $\epsilon_{ij}$ uniformly distributed and $\mu=10$, the blue curve has $\epsilon_{ij}$ uniformly distributed and $\mu=10^3$ while the black dashed curve has $\epsilon_{ij}$ normally distributed and $\mu=10^2$.}
\label{fig:fracstates}
\end{center}
 \end{figure}

In order to study this dependence, we performed a numerical computation based on finite size samples of graphs with different ansatz for the $\epsilon_{ij}$ distribution.
For each temperature $T$ we computed the fraction of states with probability no less than than $1/S$, where $S$ is the sample size. The analysis is performed by assuming $L=10^4$.
The results for two different choices of the $\epsilon_{ij}$ distribution (uniform and gaussian) and for three different values of $\mu$ are summarized in plot~\ref{fig:fracstates}. We see that our previous arguments are confirmed by these results.

Now we want to apply the same approach to real markets.For this purpose we employ data on interbank credit transactions provided by e-MID, which is the only electronic interbank market in the Euro Area and US. According to the European Central Bank, e-MID accounted, before the crisis (which is the period we take into consideration), for 17\% of total turnover in unsecured money market in the Euro Area. One distinctive feature of the platform is that it is centralized and fully transparent. Trades are public in terms of maturity, rate, volume, and time. Buy and sell proposals appear on the platform with the identity of the bank posting them, in such a way that market participants can choose their counter-parties. For more details see \cite{hol1}.

In particular, we employ data on the number of transactions between market participants, which are temporally aggregated over reserve maintenance periods. These correspond to one calendar month, i.e. about 23 trading days, which is the time frame used to compute minimum reserves requirements on the basis of the balance sheet of banks. In particular, we choose for our test the maintenance period corresponding to January 2005, for the following reasons. In this period rates were stable and characterized by a limited dispersion, since overnight lending was considered safe and asymmetric information, e.g. regarding banks' financial health, was not a significant preoccupation for market participants \cite{iori2}. Under these conditions, we cannot disregard ex ante the hypothesis that market was in equilibrium and that credit was allocated on the sole basis of supply and demand, i.e. that the interbank market could be Walrasian in the sense explained above. In fact, if credit markets are generally speaking
not centralized, with different prices (i.e. spreads) reflecting informational asymmetry and transaction costs, in this case we observe a single price (apart from small fluctuations), since interbank rates in that period were virtually identical to the policy rate. Moreover, the stability of rates over time shows that there were no significant unbalances between supply and demand, which would otherwise require an adjustment in the pricing of credit. Such a dramatic adjustment occurred indeed at the onset of the crisis, when interbank rates began diverging starkly from the policy rates, and some banks were forced to pay higher spreads if not shut out of the market.

The e-MID market in January 2005 involved 165 banks, which performed 8,705 transactions, while the number of edges connecting banks (binary links) was 2,300. In our analysis, the out- and in-strength of a bank is given by the number of borrowing or lending transactions performed in the maintenance period. The choice of not considering actual volumes of transaction (i.e. loan amounts) is a conservative one. In fact, the indifference principle underlying entropy maximization favors dense market configurations which do not correspond to the sparsity of real market data \cite{bar14,Musmeci2013,Marsili2012}, and using volumes, i.e. larger numerical values, would yield a less realistic null model.

The crucial step of our analysis is to estimate $T$ from the data by looking at the strength distributions of the real network. As we explain in the Appendix, we may identify $T$ with one of the parameters describing these distributions. Then we need to fit the out- and in-strength distributions of the network under consideration to some candidate probability density in the first place. Using the approach of \cite{clauset2009power} we test an array of alternative options by means of log-likelihood ratios, and find significant evidence in favor of the lognormal distribution (Tab. \ref{tab1})\footnote{In this comparison we set $x_{min} = 1$ in the maximum likelihood estimation of the Pareto distribution since we wish to obtain a fitted distribution for all data and not only for values the right tail.}

\begin{table}[hbt]
\centering
\begin{tabular}{@{\extracolsep{5pt}}lD{.}{.}{-3} D{.}{.}{-3} }
\hline
LLR & x & \omega \\
\hline
Lognormal \textit{vs} Pareto & 40.69^{***}&98.43^{***}\\
Lognormal \textit{vs} trunc. Pareto & 5.37&37.23^{***}\\
Lognormal \textit{vs} exponential & 31.03^{***}&8.67\\
Lognormal \textit{vs} stretched exp. & 31.03^{***}&8.67\\
\hline
\end{tabular}
\caption{Log-likelihood ratios ($^{***}$p$<$0.01)}\label{tab1}
\end{table}

\begin{figure}[htb]
\begin{center}
\subfloat[\small in-strength ($x$)]{\includegraphics[scale = 0.3]{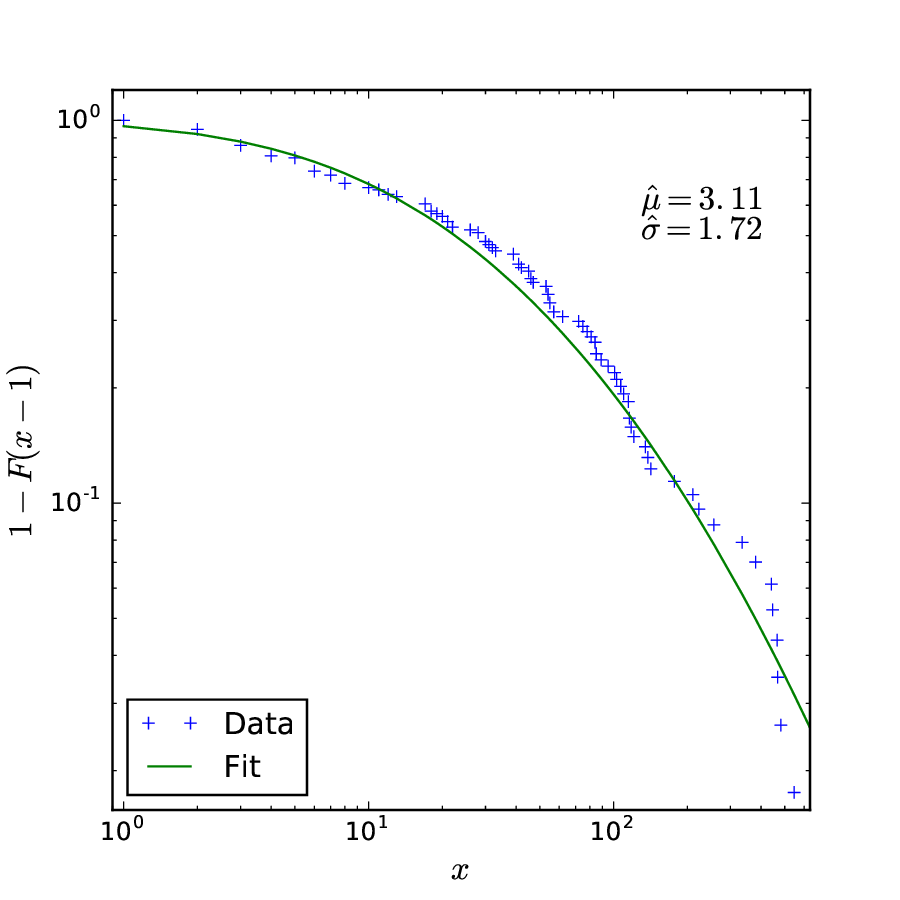}}
\subfloat[\small out-strength ($\omega$)]{\includegraphics[scale = 0.3]{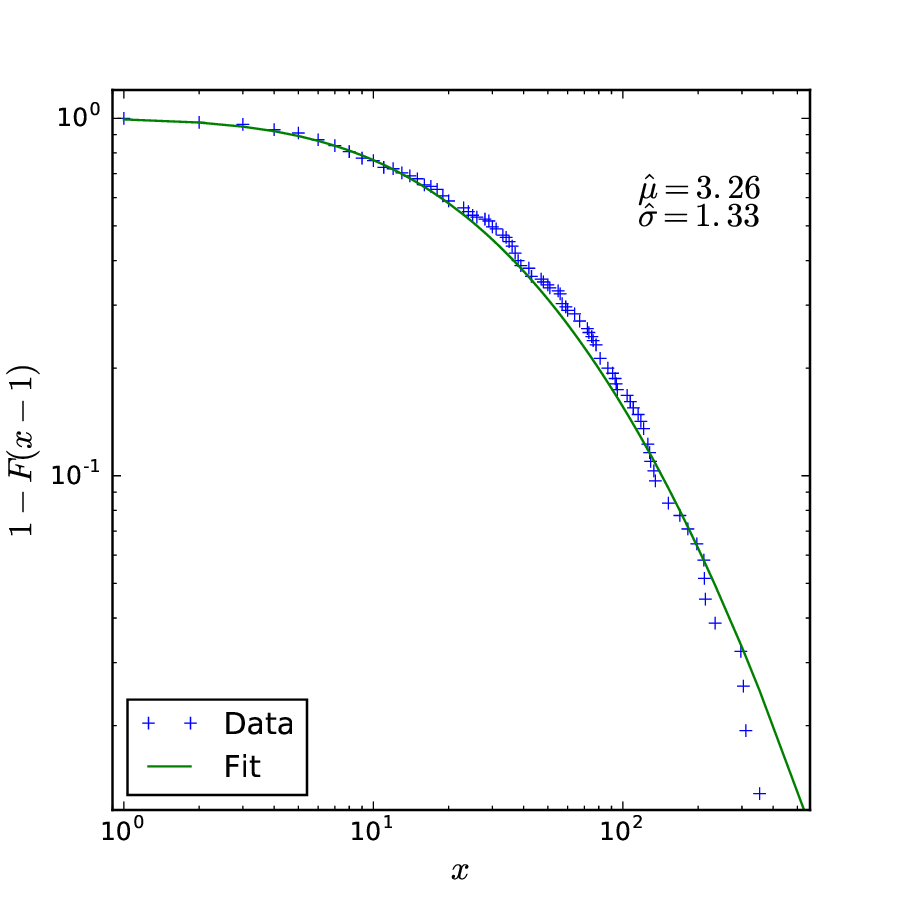}}
\end{center}
\caption{Lognormal fit of strengths, e-MID market (January 2005)} \label{fig: fit}
\end{figure}

From the fitted lognormal distributions we get two alternative estimates for $T$, namely $\sigma_{\omega}^{-1} = 0.75$ and $\sigma_{x}^{-1} = 0.58$ (see Fig. \ref{fig: fit}). We make the conservative choice of picking the larger of the two values for our exercise. In order to obtain the distribution of energies we proceed as follows. We compute the parameters $\mu$, $\lambda$ and $\rho$ introduced in eqs. (\ref{ss2})-(\ref{ss3}) using the following relationships (see Appendix):

\begin{align*}
\mu & = - T \ln L\\
\lambda_i & = - T \ln x_i   \\
\theta_i & = - T \ln \omega_i
\end{align*}

\begin{figure}[htb]
\begin{center}
\subfloat[\small $\lambda$]{\includegraphics[scale = 0.35]{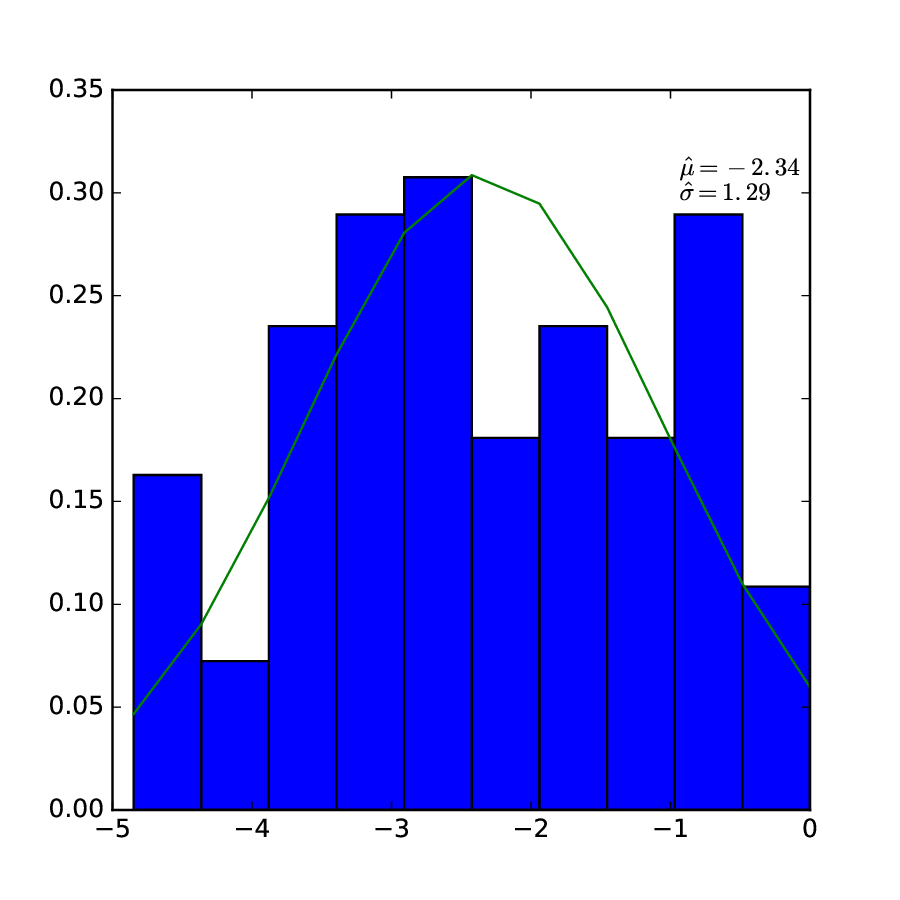}}
\subfloat[\small $\theta$]{\includegraphics[scale = 0.35]{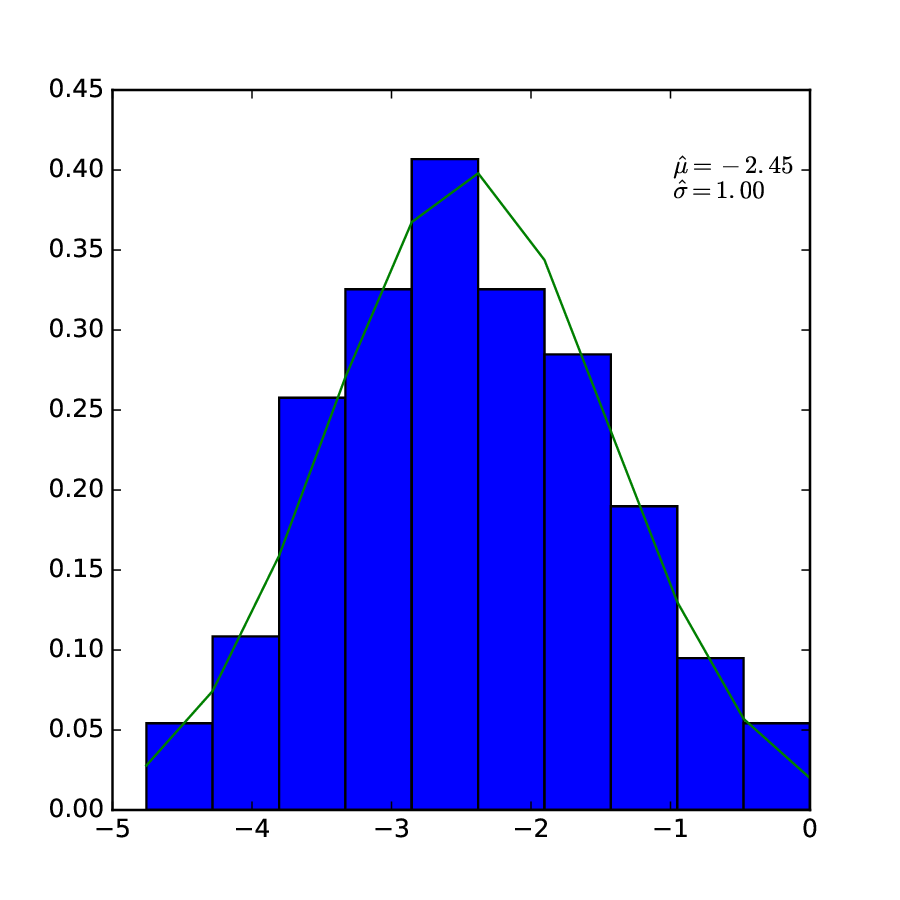}}
\subfloat[\small $\epsilon$]{\includegraphics[scale = 0.35]{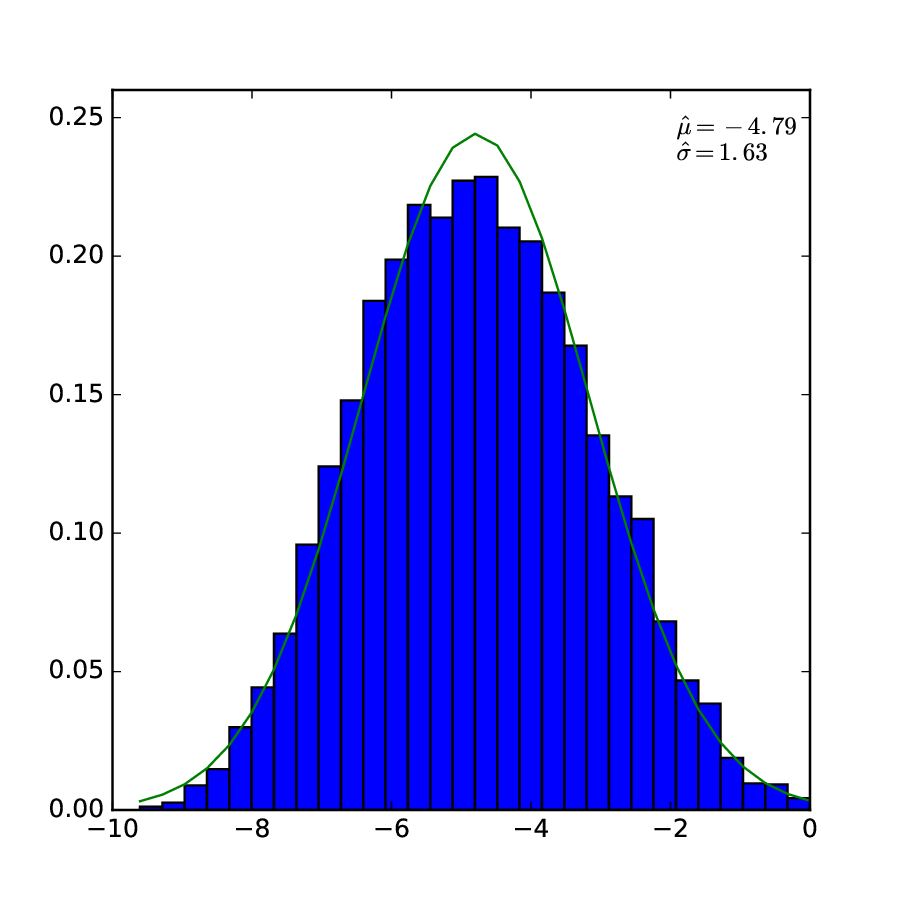}}
\end{center}
\caption{Distribution of parameters with normal fit, e-MID market (2005)} \label{fig: phi}
\end{figure}

 We immediately get $\mu = - 0.75 \times \log(8,705) =  -6.82$. Regarding $\lambda$ and $\theta$, we get two variables which are approximately normal (Fig. \ref{fig: phi}, panels (a)-(b)). The standard deviation of $\lambda$ is given by $\frac{\sigma_x}{\sigma_{\omega}} = 1.29$ as we expect from eq. (\ref{eq: normal_theta}). The distribution of the energies (panel (c)) is normal too.

\begin{figure}[ht]
\begin{center}
\includegraphics[scale=0.5]{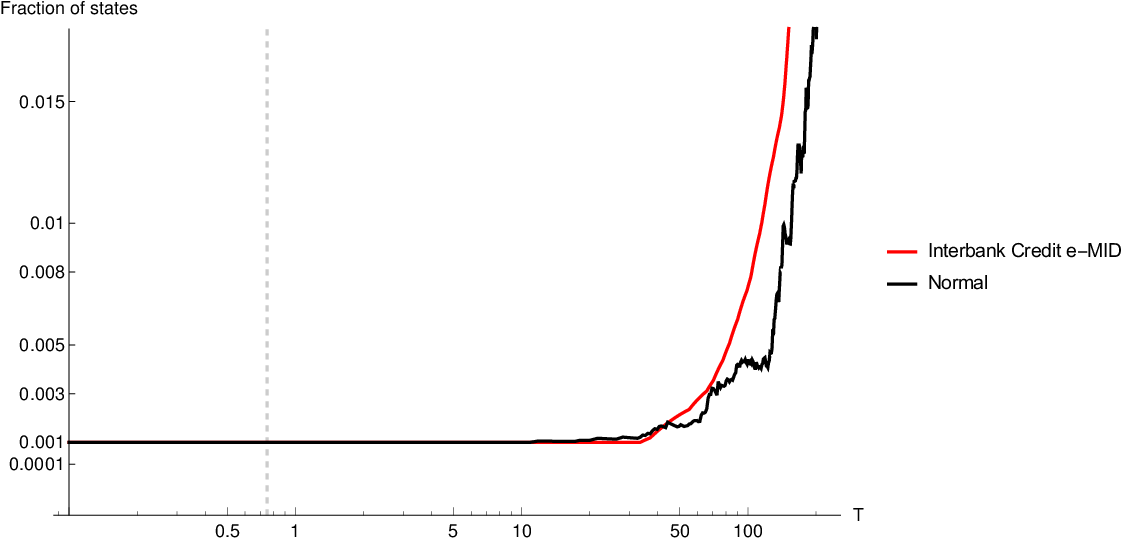}
 \caption{Fraction of states (sample size $M=10^4$) above a given probability as a function of T.
The red curve is obtained from energy level distribution of the e-MID credit market
while the black curve has $\epsilon_{ij}$ normally distributed with the same parameters of Fig. \ref{fig: phi}, panel (c). The vertical dashed line is at the estimated temperature value $T\simeq 0.75$.}
\label{fig:realfracstates}
\end{center}
\end{figure}

In Fig.~\ref{fig:realfracstates} we single out the fraction of graphs with non-negligible probability in the case of normally distributed energies with the same parameters of Fig. \ref{fig: phi}, panel (c). We see that this system tracks very closely the behavior of the e-MID market except for higher temperatures. We also see that at the estimated temperature (vertical dashed line) this fraction is extremely small, i.e. that the Walrasian model is highly predictive.

\begin{figure}[ht]
\begin{center}
\includegraphics[scale=0.5]{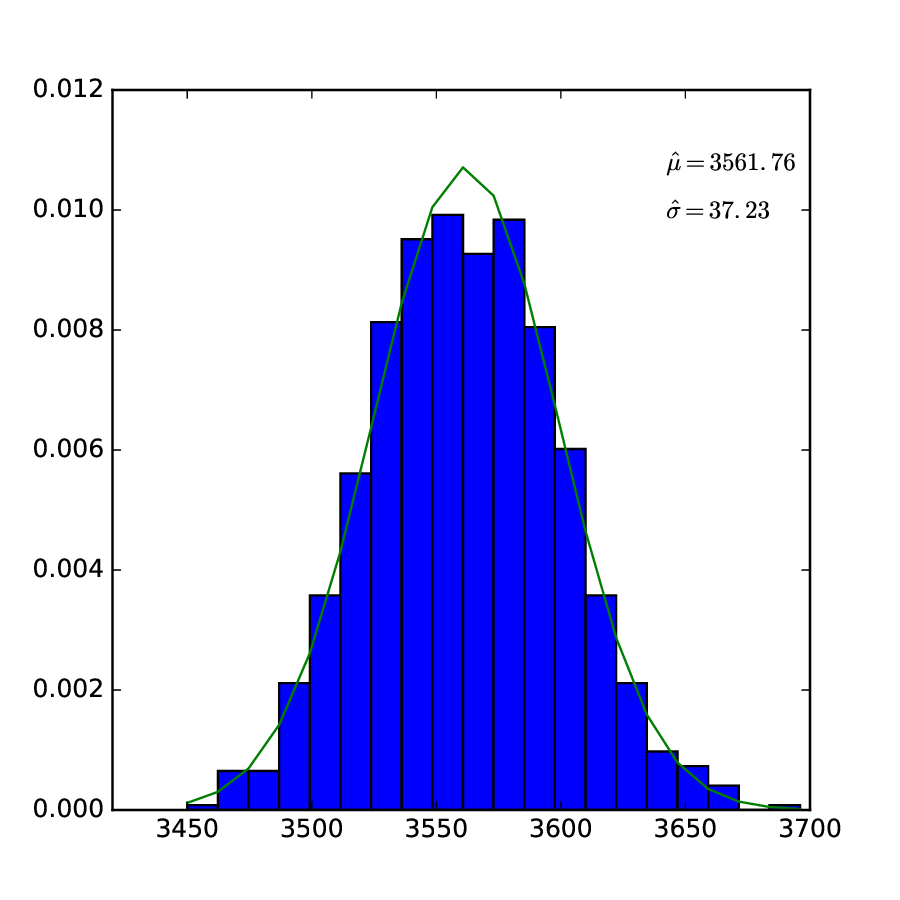}
 \caption{Distribution of $L$ in a sample of Boltzmann networks.}
\label{fig: link_dist}
\end{center}
\end{figure}

The final step is to verify if the actual market configuration $G_{\text{emid}}$ falls within the fraction of the most probable configurations. The null hypothesis we want to test is
\begin{equation}
H_0: G_{\text{emid}} \in \mathcal{G}_{WEM}
\end{equation}
In order to test this hypothesis we compare the number of binary links $l = 2,300$ observed in $G_{\text{emid}}$ with the distribution of $L$ taken from the ensemble $\mathcal{G}_{WEM}$. Since $w_{ij}$ for $G \in \mathcal{G}_{WEM}$ follows, in the Boltzmann case, a Poisson distribution with parameter $\lambda_{ij} = x_i \, \omega_j / L$, the topology of this network is given by a Bernoulli variable $a_{ij}$ with parameter $p_{ij} = 1 - \exp(-\lambda_{ij})$. We simulate a sample of binary random networks of this kind obtaining that the distribution of links is approximately normal (see Fig. \ref{fig: link_dist}).

Given the fitted parameters, the z-score of the number of links in $G_{\text{emid}}$ is -33.89.
Thus the null hypothesis is clearly rejected. This result is general, in the sense that any other market configuration with the same number of links $l$ would yield the same result. We observe that, although the expected density for $G \in \mathcal{G}_{WEM}$ is higher than the real one (0.13 \textit{vs} 0.08), the most likely network configurations in this ensemble do not display a complete topology. 

\section{Discussion}\label{sec: disc}

Samuelson expressed the view that no more than a formal mathematical analogy exists between classical thermodynamics and utility theory\cite{samuel60}. This analogy is grounded in the mathematical tool of constrained optimization, which is applied respectively to entropy and to the individual utility function.
More recently, other authors have pushed further this parallel \cite{3b}.  It is interesting to compare their approach with the one we present in this paper. From a methodological point of view, they are quite different: the former leverages on the parallel between utility and entropy, and consequently tries to mutually adapt neoclassical utility theory and classical thermodynamics; the latter leverages instead on the indeterminacy of WE with respect to market configurations in order to bring statistical physics into general equilibrium theory, and then thermodynamics is derived in the usual way of physics.\\
Our approach is more similar to the one of \cite{foley94}, advocating a statistical equilibrium theory of markets. There are however some essential differences. In \cite{foley94} the maximum entropy principle is employed to introduce a different notion of market equilibrium with respect to WE. In this framework economic agents do not necessarily choose their optimal allocation, instead they randomly choose among a set $Z$ of alternative, welfare improving, net demand vectors of $m$ commodities. The maximum entropy principle is employed in order to derive the distribution of agents over the net demand set $Z$.
The net demand vectors of \cite{foley94}, once summed with endowments, represents the final allocations after individual market transactions took place. In this paper we focus instead on the maximum entropy distribution of transactions themselves, conditioned to a given equilibrium allocation. Such description is possible also for statistical equilibria in the sense of \cite{foley94}, since the latter assign a unique final allocation to each agent. \\
The generalization of our framework to multiple markets can be achieved introducing the concept of multi-level or multiplex networks \cite{bianconi13}. In particular, we can treat the $m$ markets as a set of statistically independent networks each satisfying the allocations of \cite{foley94} which now take the place of Walrasian optimal allocations. 
All the arguments of sec. \ref{sec: appl} still apply, and we come to see that this weaker economic equilibrium is no less restrictive than WE with respect to market configurations. In particular, it still favors dense over sparse configurations. Thus our results apply not only to WE, but even to weaker equilibria, as long as final allocations are the only variables governing transactions.

\section{Conclusions}

We started in Sec. 3 from general equilibrium theory by introducing the equiprobable WE states as the building blocks of a statistical representation of market configurations. This provides a natural link with complex networks, and in particular a clear connection between graph and thermodynamic variables which preserves the core of microeconomic theory.
As an application, in Sec. 4 we have considered 
different simple expressions for the ``energies''  $\epsilon_{ij}$,
and studied the ensemble probability of a given market configuration in terms of the temperature $T$.
Moreover,  we compared the behavior
of a real market (e-MID interbank credit market)
with an artificial system with normally distributed energies. As a result, we found that
the number of states with non negligible probability tracks closely the normally distributed case except for higher temperatures. We also found that at the estimated temperature the fraction of states with non negligible probability is minimal.

The main lesson of our approach is that, in order to characterize economic equilibrium, only a small number of market configuration must be considered. This makes WE a highly restrictive theory with respect to its empirical implications. In fact, we rejected the null hypothesis that the observed e-MID credit market configuration is a WE.
We arrive thus at the conclusion that WE is empirically testable and highly restrictive with respect to market configurations. Instead, according to the standard interpretation of the well known Sonnenschein-Mantel-Debreu theorem, WE cannot be falsified by the empirical evidence provided by price effects. 
 
The intuition behind our results is the following.
Node variables, like strength or degree,
cannot in general explain all the properties of real market configurations. This observation is consistent with a recent stream of network models tackling the problem of real networks' estimation \cite{3g,bar14,Musmeci2013,Marsili2012}. For instance, interbank markets display non trivial higher order topological properties, like high reciprocity, low clustering \cite{squartini2013}, community
structure \cite{bargall2013}, strength-degree correlations \cite{Musmeci2013}, and preferential relationships \cite{iori2}, which cannot be derived from the weighted configuration model which is associated with WE. That's why other variables, in addition to allocations which are the only relevant variables in WE, must be introduced in order to obtain a realistic theory of markets.

\vspace{12pt}

\noindent \begin{footnotesize}\textbf{Acknowledgments:} we are deeply thankful to Giulia Iori for providing us with the e-MID data. 
We thank the editor and one anonymous reviewer for their 
constructive comments, which helped us to improve the manuscript All the usual disclaimers apply.\end{footnotesize}


\section*{Appendix}
In this appendix we show how to compute the grandcanonical partition function for the network equivalents of the three ideal gas systems. Let's start by computing $\mathcal{Q}$ in the case of a directed Fermionic (i.e. binary) network.
Since $L=\sum_{i\neq j}{a}_{ij}$,
the grand partition function reads (recall $\beta = \frac{1}{T}$):
\begin{align}
\mathcal{Q}(z,V,T) & = \sum_{L = 0}^{n(n-1)} z^L Z_G(L,V,T) = 
\sum_{\{{a}_{ij}\}} e^{\beta(\mu L-H)} \label{ggg}  \\
& = \sum_{\{{a}_{ij}\}}
e^{\sum_{i\neq j}\beta(\mu - \epsilon_{ij}){a}_{ij}}= \\
& = \prod_{i \neq j} \left [ 1 + e^{\beta(\mu - \epsilon_{ij})}\right]. \label{gg1}
\end{align}
The expected occupation numbers $\langle {a}_{ij} \rangle$ are obtained in the usual way
\begin{align}
\langle {a}_{ij} \rangle & = - \frac{1}{\beta} \dfrac{\partial}{\partial \epsilon_{ij}} \log \mathcal{Q}\\
& = \dfrac{e^{\beta(\mu - \epsilon_{ij})}}{1 + e^{\beta(\mu - \epsilon_{ij})}},
\end{align}
which coincides with the results of Park \& Newman for $T = 1$. In the Bosonic (weighted) case, instead of (\ref{gg1}) we have
\begin{equation}
\mathcal{Q}(z,V,T)=
\prod_{i \neq j} \left [\frac{1}{1 - e^{\beta(\mu - \epsilon_{ij})}} \right].
\label{gg2}
\end{equation}
The expression (\ref{gg2}) gives
\begin{equation}
\langle {w}_{ij} \rangle=
\frac{e^{\beta(\mu - \epsilon_{ij})}}{1 - e^{\beta(\mu - \epsilon_{ij})}},
\label{gg3}
\end{equation}
that is the Bose-Einstein distribution for networks. Finally for a Boltzmann (weighted) network, taking into account that in this case 
$W(G) = \prod_{i \neq j} w_{ij}!$, we have \cite{bar14}:
\begin{equation}
\mathcal{Q}(z,V,T)=
\prod_{i \neq j} \exp \left [e^{\beta(\mu - \epsilon_{ij})}\right].
\end{equation}
and
\begin{equation}
\langle {w}_{ij} \rangle = e^{\beta(\mu - \epsilon_{ij})}
\end{equation}
We remark that the network analogue of the Boltzmann gas, which implies that the ${w}_{ij}$ are Poisson variables, is generally overlooked in the literature or introduced
with ad hoc assumptions \cite{karrer2011}. It's easy to see that, with the constraints (\ref{eq: constr1})-(\ref{eq: constr2}), in this case we arrive at the following explicit ME solution \cite{bar2011396}:
\begin{equation} \label{eq: expl_boltz}
 \left \langle w_{ij} \right \rangle = \frac{\omega_ix_j}{L} \qquad \qquad i,j = 1, \dots, N
\end{equation}
From this solution we can obtain the values for the parameters $\mu , \lambda_i, \theta_i$:
\begin{align*}
\mu & = - T \ln L\\
\lambda_i & = - T \ln x_i \\
\theta_i & = - T \ln \omega_i  
\end{align*}
Following \cite{2006cond.mat..6805G} we can thus connect the probability distribution of these parameters with the probability distribution of the corresponding observables, namely strengths. We also observe that nodes with the same strength values are also characterized by the same parameter values. We can derive the same results for Bosonic and Fermionic systems by observing that
\begin{align}
\frac{\partial \langle a_{ij} \rangle}{\partial \lambda_i} & = \frac{\partial \langle a_{ij} \rangle}{\partial \theta_j} = \beta \, \langle a_{ij} \rangle \, \left[ \langle a_{ij} \rangle - 1 \right ] < 0 \\
\frac{\partial \langle w_{ij} \rangle}{\partial \lambda_i} & = \frac{\partial \langle w_{ij} \rangle}{\partial \theta_j} = - \beta \, \langle w_{ij} \rangle \, \left[ \langle w_{ij} \rangle + 1 \right ] < 0
\end{align}
Then from 
\begin{equation} \label{eq: str_equal}
\sum_j  \frac {e^{\beta(\mu - \lambda_{i} - \theta_{j})}}{1 \pm e^{\beta(\mu - \lambda_{i} - \theta_{j})}} = \sum_j \frac {e^{\beta(\mu - \lambda_{h} - \theta_{j})} }{1 \pm e^{\beta(\mu - \lambda_{h} - \theta_{j})}}
\end{equation}
we see that if $\lambda_i \neq \lambda_h$ then the equality (\ref{eq: str_equal}) cannot be satisfied.  Unfortunately, in the Bosonic and Fermionic cases we cannot write down explicitly the functional dependency between strengths and parameters. In the Boltzmann case, instead, we can derive the distribution of the parameters from the distribution of the constraints, at least in some cases. For this reason in the following and in sec. \ref{sec: appl} we always refer to Boltzmann systems. 
The authors of \cite{2006cond.mat..6805G}  showed that, if $ \rho(w)  = (\gamma - 1) \, w^{-\gamma} \, dw$ then $\phi = T \ln w$ is exponentially distributed in a Boltzmann system:
\begin{equation}
\rho(\phi) =  \frac{\gamma - 1}{T}\,\exp \left (-\phi\frac{\gamma - 1}{T} \right ) \, d\phi
\end{equation}
They further observed that we can make $\rho(\phi)$ independent of $T$ by setting $T = \gamma - 1$. In this way we obtain that the probability density of $\phi$ regains some generality, in the sense that, as long as $\rho(w)$ is in the family of Pareto distributions, $\rho(\phi)$ is unaffected by variations of $T$. Then we can envisage a simple procedure to estimate $T$ from $\rho(w)$, since we can estimate $\gamma$ from the data. This argument requires that we take a continuous approximation of $\rho(w)$ in case that the strengths are integers. We can extend it to other densities, like for instance $\rho(w) = \frac{1}{w \sigma \sqrt{2 \pi}} \, \exp\left( - \frac{(\ln w)^2}{2\sigma^2}\right) \,dw$. In this case, by operating a substitution of variables we obtain
\begin{equation}
\rho(\phi) = \frac{1}{\sigma\,T\,\sqrt{2\pi}}\,\exp\left(-\frac{\phi^2}{2\,(\sigma T)^2} \right)\,d\phi
\end{equation}
By setting $T = \sigma^{-1}$ we obtain that $\phi$ is distributed like a standard normal variable.
The author of \cite{2006cond.mat..6805G} develop their argument in the symmetric case. In the asymmetric / directed case we are confronted with two strength distributions $\rho(\omega)$ and $\rho(x)$. Supposing that both come from the same family (Pareto) but with different parameters (respectively $\gamma_{\omega}$ and $\gamma_{x}$), we can set $\rho(\lambda) = \exp(\lambda)$ (since $\lambda \leqslant 0$) with $T = \gamma_{x} - 1$. Then we obtain ($\theta \leqslant 0$)
\begin{align}
\rho(\theta) & = \frac{\gamma_{\omega} - 1}{T} \exp \left ( \theta\,\frac{\gamma_{\omega} - 1}{T} \right ) \,d\theta \\
& = \frac{\gamma_{\omega} - 1}{\gamma_{x} - 1} \exp \left ( \theta\,\frac{\gamma_{\omega} - 1}{\gamma_{x} - 1} \right )\,d\theta
\end{align}
In the lognormal case, supposing that $T = \sigma_{x}^{-1}$, we obtain that
\begin{align}
\rho(\lambda) & = \frac{1}{\sqrt{2\pi}}\,\exp\left(\frac{-\lambda^2}{2\,} \right)\,d\lambda \\
\rho(\theta) & =  \frac{1}{\sigma_{\omega}\sigma_{x}^{-1}\sqrt{2\pi}}\,\exp \left(\frac{-\theta^2}{2\,\sigma_{\omega}\sigma_{x}^{-1}} \right)\,d\theta \label{eq: normal_theta}
\end{align}

\end{document}